\documentclass[conference,10pt]{IEEEtran}
\usepackage{diagbox} 
\usepackage{graphicx}
\usepackage{epstopdf}
\usepackage{psfrag}
\usepackage{subfigure}
\usepackage{url}
\usepackage{stfloats}
\usepackage{amsfonts,amssymb,amsmath,bm,paralist,theorem,cite,ifthen,color,nccmath}
\usepackage{caption}
\usepackage{calc}
\usepackage{enumerate}
\usepackage{multirow}
\usepackage[ruled]{algorithm2e}
\usepackage{array}
\usepackage{float}
\usepackage{soul}
\newtheorem{proposition}{Proposition}

\graphicspath{{Figures/}}

\newcommand{\T}{{\scriptscriptstyle\mathsf{T}}}
\renewcommand{\H}{{\scriptscriptstyle\mathsf{H}}}

\newcommand\Ccl{\ensuremath{\mathcal{C}}}
\newcommand\Dcl{\ensuremath{\mathcal{D}}}

\newcommand\Ncl{\ensuremath{\mathcal{N}}}
\newcommand\Ocl{\ensuremath{\mathcal{O}}}

\newcommand\Tcl{\ensuremath{\mathcal{T}}}

\newcommand\Ucl{\ensuremath{\mathcal{U}}}

\newcommand\Cs{\ensuremath{{\mathbb{C}}}}
\newcommand\Es{\ensuremath{{\mathbb{E}}}}
\newcommand\Rs{\ensuremath{{\mathbb{R}}}}

\newcommand\Ab{\ensuremath{ \mathbf{A} }}

\newcommand\Cb{\ensuremath{ \mathbf{C} }}

\newcommand\Hb{\ensuremath{ \mathbf{H} }}
\newcommand\Ib{\ensuremath{ \mathbf{I} }}

\newcommand\Ub{\ensuremath{ \mathbf{U} }}

\newcommand\Vb{\ensuremath{ \mathbf{V} }}

\newcommand\Jb{\ensuremath{ \mathbf{J} }}

\newcommand\Xb{\ensuremath{ \mathbf{X} }}
\newcommand\Yb{\ensuremath{ \mathbf{Y} }}
\newcommand\Zb{\ensuremath{ \mathbf{Z} }}

\newcommand\ab{\ensuremath{ \mathbf{a} }}

\newcommand\eb{\ensuremath{ \mathbf{e} }}

\newcommand\hb{\ensuremath{ \mathbf{h} }}

\newcommand\qb{\ensuremath{ \mathbf{q} }}
\newcommand\ssb{\ensuremath{ \mathbf{s} }}
\newcommand\rb{\ensuremath{ \mathbf{r} }}
\newcommand\tb{\ensuremath{ \mathbf{t} }}
\newcommand\ub{\ensuremath{ \mathbf{u} }}
\newcommand\vb{\ensuremath{ \mathbf{v} }}
\newcommand\wb{\ensuremath{ \mathbf{w} }}

\newcommand\zb{\ensuremath{ \mathbf{z} }}

\newcommand\Sigmab{\ensuremath{{\bm \Sigma}}}

\newcommand\etab{\ensuremath{{\bm \eta}}}

\newcommand\diag{\ensuremath{{\rm diag}}}
\newcommand\tr{\ensuremath{{\rm Tr}}}

\newcommand{\SNR}{\textrm{SNR}}
\newcommand\ghz{\textrm{GHz}}

\newcommand\dB{\textrm{dB}}
\newcommand\mW{\textrm{mW}}

\newcommand\Ubb{\ensuremath{ \mathbf{U}_{\rm BB} }}
\newcommand\Urf{\ensuremath{ \mathbf{U}_{\rm RF} }}
\newcommand\Frf{\ensuremath{ \mathbf{F}_{\rm RF} }}

\newcommand\Urfbar{\ensuremath{ \bar{\mathbf{U}}_{\rm RF} }}

\newcommand\Nrf{\ensuremath{ N_{\rm RF} }}

\newcommand\Nc{\ensuremath{ N_{\rm c} }}

\newcommand\Tu{\ensuremath{ T_{\rm u} }}

\newcommand\fs{\ensuremath{ f_{\rm s} }}
\newcommand\fc{\ensuremath{ f_{\rm c} }}

\newcommand\Bw{\ensuremath{ B_{\rm w} }}

\newcommand\sigman{\ensuremath{ \sigma_{\rm n} }}

\usepackage[labelsep=period,font=footnotesize]{caption}

\usepackage[top=0.75in, bottom=1.1in, left=0.625in, right=0.625in]{geometry}

\IEEEoverridecommandlockouts

\begin{document}

\title{Hybrid Receiver Design for Massive MIMO-OFDM with Low-Resolution ADCs and Oversampling}
\author{
\IEEEauthorblockN{Mengyuan Ma, Nhan Thanh Nguyen, Italo Atzeni, and Markku Juntti}
\IEEEauthorblockA{Centre for Wireless Communications (CWC), University of Oulu, Finland \\
E-mail: \{mengyuan.ma, nhan.nguyen, italo.atzeni, markku.juntti\}@oulu.fi
}
}

\maketitle

\begin{abstract}
    Low-resolution analog-to-digital converters (ADCs) and hybrid beamforming have emerged as efficient solutions to reduce power consumption with satisfactory spectral efficiency (SE) in massive multiple-input multiple-output (MIMO) systems. In this paper, we investigate the performance of a hybrid receiver in massive MIMO orthogonal frequency-division multiplexing (OFDM) uplink systems with low-resolution ADCs and oversampling. Considering both the temporal and spatial correlation of the quantization distortion (QD), we derive a closed-form approximation of the frequency-domain QD covariance matrix, which facilitates the evaluation of the system's SE. Then we jointly design the analog and digital combiners of the hybrid receiver to maximize the SE. The formulated problem is challenging due to the constant-modulus constraint of the analog combiner and its coupling with the digital one. To overcome these challenges, we transform the objective function into an equivalent but more tractable form and then iteratively update the analog and digital combiners. Numerical simulations verify the superiority of the proposed algorithm over the considered benchmarks and show the resilience of the hybrid receiver to beam squint with low-resolution ADCs. Furthermore, the proposed hybrid receiver design with oversampling can achieve significantly higher energy efficiency compared with the fully digital one.
\end{abstract}
	

\section{Introduction}

Massive multiple-input multiple-output (MIMO) technology stands as a cornerstone in current and future wireless systems \cite{Raj20}, leveraging large antenna arrays to deliver exceptional spectral efficiency (SE) \cite{albreem2019massive}. However, the deployment of massive MIMO at millimeter-wave and \mbox{(sub-)THz} frequencies presents a significant challenge: the high power consumption associated with the numerous radio-frequency (RF) chains poses a threat to the system's energy efficiency (EE) \cite{bjornson2019massive,shafie2022terahertz}. On the one hand, the analog-to-digital converters (ADCs) are typically the most power-hungry RF components, as their power consumption increases exponentially with the number of resolution bits \cite{murmann2015race}. For instance, high-speed ADCs with 8--12 bits operating at 1~Gsample/s can consume several Watts \cite{li2017channel}, which motivates the adoption of low-resolution ADCs. On the other hand, hybrid analog-digital beamforming techniques can effectively reduce the required number of RF chains without unduly compromising the SE \cite{ma2021closed}. In this context, the integration of low-resolution ADCs within hybrid beamforming architectures emerges as a compelling strategy to balance energy cost and SE performance \cite{Atz22,abbas2017millimeter}.

The performance of hybrid receivers with low-resolution ADCs was investigated in \cite{abbas2017millimeter,mo2017hybrid,roth2018comparison,ma2024digital}, demonstrating a flexible trade-off between power consumption and SE performance. However, these works mainly focus on narrowband single-carrier systems and, thus, their results are not applicable to wideband multi-carrier systems, e.g., operating in the \mbox{(sub-)THz} band. This is because the analog combiner in the hybrid receiver is frequency-flat and must be designed for all the subcarriers. Furthermore, the coarse quantization introduces additional correlation of the time-domain symbols, which affects the frequency-domain performance analysis \cite{jacobsson2019linear}. The performance of MIMO orthogonal frequency-division multiplexing (OFDM) uplink systems with low-resolution ADCs was investigated in \cite{mollen2016uplink,studer2016quantized,uccuncu2019performance,ma2023analysis}. For example, it was shown in \cite{mollen2016uplink,studer2016quantized} that few-bit quantization in MIMO-OFDM systems can approach the performance of the full-resolution case. 
In our previous work \cite{ma2023analysis}, we showed that oversampling in a MIMO-OFDM uplink system can significantly enhance the sum rate, especially at high signal-to-noise ratios (SNRs) and with very low ADC resolutions. However, \cite{mollen2016uplink,studer2016quantized,uccuncu2019performance,ma2023analysis} only investigated fully digital receivers, leaving the hybrid receiver design for low-resolution wideband systems unexplored. Furthermore, while the beam squint effect has been shown to significantly degrade the performance of hybrid beamforming with full-resolution ADCs \cite{dai2022delay,ma2023beam,ma2024switch}, its impact on low-resolution systems remains an open question. 

To bridge the knowledge gap described above, we herein investigate a massive MIMO-OFDM uplink system employing both low-resolution ADCs and hybrid beamforming. Specifically, we first derive a closed-form approximation for the frequency-domain covariance matrix of the quantization distortion (QD), which considers both the temporal and spatial correlation. With this approximation, we jointly design the analog and digital combiners to maximize the SE. The formulated problem is complicated by the constant-modulus constraint of the analog combiner and its coupling with the digital one. To overcome these challenges, we transform the objective function into an equivalent but more tractable form and then iteratively update the analog and digital combiners. Numerical simulations verify the superiority of the proposed algorithm over the considered benchmarks. Furthermore, the results demonstrate that the performance degradation of the hybrid receiver due to the beam squint effect is less significant with low-resolution ADCs. Lastly, the EE comparison shows that the proposed hybrid receiver design with oversampling significantly outperforms the considered fully digital one.

\section{System and Quantization Models}

\subsection{Signal Model}

We consider an uplink massive MIMO system where a base station (BS) equipped with $M$ antennas receives signals from $I$ single-antenna user equipments (UEs). We assume that the BS deploys a fully connected hybrid beamforming architecture in which each RF chain connects to all the antennas through a phase-shifter network. Let $\Nrf$ be the number of RF chains at the BS. OFDM is assumed over a wideband channel to deal with the frequency selectivity. Specifically, let $\Delta f= \frac{1}{\Tu}$ be the subcarrier spacing, where $\Tu$ denotes the OFDM symbol duration, which is assumed to be fixed. Let $f_{k}=\fc+\left(k+1-\frac{\Nc+1}{2} \right)\Delta f$, $k=0,\ldots, \Nc-1$, denote the $k$-th subcarrier frequency, where $\fc$ is the center carrier frequency. Among the $\Nc$ subcarriers, $K$ of them are dedicated to the signal transmission while the other $\Nc-K$ are employed for oversampling \cite{jacobsson2019linear}. Let $\tilde{s}_i[k]$ be the transmit symbol of the $i$-th UE at the $k$-th subcarrier, with $\Es\left[|\tilde{s}_i[k]|^2\right]=1$ for $k=0,\ldots, K-1$. Note that, when $\Nc >K$, we have $\tilde{s}_i[k]=0$ for $k=K,\ldots,\Nc-1 $. Since the sampling frequency is $\fs=\Nc \Delta f$ and the transmission bandwidth is $\Bw=K \Delta f$, the oversampling ratio (OSR) is defined as $\beta= \frac{\Nc}{K}$. We assume that $\Nc$ can be adjusted, while $K$ is fixed. Hence, $\beta=1$ and $\beta>1$ indicate the Nyquist sampling and the oversampling scheme, respectively. The time-domain symbol is obtained by an $\Nc$-point inverse discrete Fourier transform (IDFT), i.e.,
\begin{equation}\label{eq:IDFT of stt frequency}
    s_i[n]=\sqrt{\frac{p}{\Nc}}\sum\limits_{k=0}^{\Nc-1}\tilde{s}_i[k] e^{j\frac{2\pi nk}{\Nc}}, \quad n=0,\ldots, \Nc-1,
\end{equation}
where $n$ represents the index of the time-domain symbol and $p$ denotes the average transmit power for the symbols. At the receiver, the time-domain signals are first downconverted to the digital domain and then transformed back to the frequency domain by an $\Nc$-point discrete Fourier transform (DFT).

Define $\ssb[n]=\left[ s_1[n], \ldots,s_{I}[n] \right]^\T$ and let $\Hb[d]=\left[\hb_{1}[d],\cdots,\hb_{I}[d] \right]\in \Cs^{M\times I}$ denote the channel matrix at the $d$-th delay tap, where $\hb_i[d]$ represents the channel of the $i$-th UE. The discrete-time received signal at the BS at time sample $n$ is given by
 \vspace{-2mm}
\begin{equation}\label{eq:compact form at receiver}
	\rb[n]=\sum_{d=0}^{D-1} \Hb[d] \ssb[n-d] +\wb[n],
\end{equation}
with $D=\beta D_0$, where $D_0$ is the maximum number of delay taps under Nyquist sampling and $ \wb[n] \sim \Ccl\Ncl(\mathbf{0},\sigma^2_{\rm n}\Ib_{M})$ represents the additive white Gaussian noise (AWGN) vector with variance $\sigman^2$. Now, define $\tilde{\ssb}[k]=[\tilde{s}_1[k], \ldots, \tilde{s}_{I}[k] ]^\T$, where $\tilde{s}_{i}[k]$ is the frequency-domain signal transmitted by the $i$-th UE, and $\tilde{\Hb}[k]=[\tilde{\hb}_1[k],\ldots,\tilde{\hb}_I[k] ]$, with $\tilde{\hb}_i[k]=\sum_{d=0}^{D-1}\hb_i[d] e^{-j\frac{2\pi d k}{N_{\rm c}}}$. By taking the DFT of both sides of \eqref{eq:compact form at receiver}, the frequency-domain received signal is expressed as
\begin{equation}
	\tilde{\rb}[k] = \sqrt{p} \tilde{\Hb}[k] \tilde{\ssb}[k]+\tilde{\wb}[k], \quad k=0, \ldots, \Nc-1, \label{eq:frequency-domain unquantized model}
\end{equation}
with $\tilde{\rb}[k]  =\frac{1}{\sqrt{\Nc}} \sum_{n=0}^{N_{\rm c}-1}\rb[n] e^{-j\frac{2\pi n k}{N_{\rm c}}}$ and $\tilde{\wb}[k]=\frac{1}{\sqrt{\Nc}}\sum_{n=0}^{N_{\rm c}-1}\wb[n] e^{-j\frac{2\pi nk}{N_{\rm c}}}$. Note that $\tilde{\rb}[k]=\tilde{\wb}[k]$ for $k=K,\ldots,\Nc-1$ as $\tilde{\ssb}[k]=\mathbf{0}$ in these cases.

\subsection{Quantization Model}


We assume that the ADCs at all the RF chains are scalar quantizers and have the same number of resolution bits $b$. Define the codebook of a quantizer of $b$~bits as $\Ccl=\{c_0, \ldots, c_{N_{\rm q}-1}\}$, where $N_{\rm q}=2^b$ is the number of output levels. The set of quantization thresholds is $\Tcl=\{ t_0, \ldots, t_{N_{\rm q}} \}$, where $t_0=-\infty$ and $t_{N_{\rm q}}=\infty$ allow inputs with arbitrary power. For standard Gaussian signals, the Lloyd-Max algorithm \cite{max1960quantizing} can find optimal sets $\Ccl$ and $\Tcl$ that achieve the minimum mean squared error (MMSE) between the input and output of the quantizer. 
Let $Q(\cdot)$ denote the quantization function. For a complex signal $x=\Re\{x\}+j \Im\{x\}$, we have
$Q(x)=Q(\Re\{x\})+j Q(\Im\{x\})$ with $Q(\Re\{x\})=c_{i}$ when $\Re\{x\}\in [t_i, t_{i+1}]$, and $Q(\Im\{x\})$ can be obtained in a similar fashion. When the input signal of the quantizer is a vector, $Q(\cdot)$ is applied elementwise.

 To model the quantization of the received signal in \eqref{eq:compact form at receiver}, we first rewrite \eqref{eq:compact form at receiver} as
\begin{equation}\label{eq: signal model time domain compact}
	\Bar{\rb}= \left[\rb[N_{\rm c}-1]^\T, \ldots, \rb[0]^\T \right]^\T = \Bar{\Hb}\Bar{\ssb} + \bar{\wb},
\end{equation}
with $\Bar{\ssb}=[\ssb[N_{\rm c}-1]^\T, \ldots, \ssb[0]^\T]^\T$ and $\Bar{\wb}=\left[\wb[N_{\rm c}-1]^\T, \ldots, \wb[0]^\T \right]^\T$, and where $\Bar{\Hb} \in \mathbb{C}^{M N_{\rm c} \times IN_{\rm c}}$ is a block circulant matrix \cite{uccuncu2019performance}. Before quantization, the received signal is first combined in the analog domain. Denote the analog combiner as $\Urf\in \Cs^{M\times \Nrf}$, which has constant-modulus entries. At the BS, the received time-domain signal $\Bar{\rb}$ is first combined with $\Urf$ and then fed into the ADCs. Therefore, the output of the ADCs is given by
\begin{align}\label{eq:quantization model}
    \Bar{\zb}=Q(\bar{\mathbf{U}}_{\rm RF}^\H \Bar{\rb}),
\end{align}
with $\Urfbar = \Ib_{\Nc}\otimes \Urf$, where $\otimes$ denotes the Kronecker product. The Bussgang-based additive quantization noise model (BAQNM) allows to model a nonlinear input-output relation of a Gaussian signal as a linear transformation \cite{ma2024joint}. With the BAQNM, \eqref{eq:quantization model} can be expressed as
\begin{equation}\label{eq:ADC model}
	\Bar{\zb}= \left[\zb[N_{\rm c}-1]^\T, \ldots, \zb[0]^\T \right]^\T = \alpha \bar{\mathbf{U}}_{\rm RF}^\H \Bar{\rb} + \Bar{\etab},
\end{equation}
where $\Bar{\etab}=\left[\etab[N_{\rm c}-1]^\T, \ldots, \etab[0]^\T \right]^\T$ denotes the non-Gaussian QD vector that is uncorrelated with $\Bar{\rb}$, $\alpha=1-\gamma$ represents the Bussgang gain, and $\gamma$ denotes the distortion factor of the ADCs, which depends on the number of resolution bits as $\gamma(b) \approx 2^{-1.74b+0.28}$ \cite{ma2024joint}. Furthermore, \eqref{eq:ADC model} is equivalent to
\begin{equation}\label{eq:quantization model for subcarriers}
		\zb[n]=\alpha \Urf^\H \rb[n] + \etab[n], \quad n=0,\ldots,\Nc-1.
\end{equation}
By taking the DFT of both sides of \eqref{eq:quantization model for subcarriers}, we have
\begin{align}
	\tilde{\zb}[k]& = \frac{1}{\sqrt{\Nc}}\sum_{n=0}^{N_{\rm c}-1}\zb[n] e^{-j\frac{2\pi nk}{N_{\rm c}}} =\alpha \Urf^\H  \tilde{\rb}[k] + \tilde{\etab}[k] \nonumber \\
    &=\alpha \sqrt{p} \Urf^\H  \tilde{\Hb}[k] \tilde{\ssb}[k]+\eb[k], \quad k=0, \ldots, \Nc-1,
\end{align}
with $\tilde{\etab}[k] =\frac{1}{\sqrt{\Nc}}\sum_{n=0}^{N_{\rm c}-1}\etab[n] e^{-j\frac{2\pi n k}{N_{\rm c}}} $ and where $\eb[k]=\alpha \Urf^\H \tilde{\wb}[k]+ \tilde{\etab}[k]$ includes both the AWGN and QD. Note that the latter is non-Gaussian due to the presence of $\tilde{\etab}[k]$.

Let $\ub_i[k]\in \Cs^{\Nrf}$ denote the digital combining vector of the $i$-th UE at the $k$-th subcarrier and define $\Ubb[k]= \left[\ub_1[k],\ldots,\ub_{I}[k]\right]$. 
Finally, the post-combined signal of the $i$-th UE at the $k$-th subcarrier is given by
	\begin{align}\label{eq:combined signal model}
	   \hat{x}_i[k]&= \underbrace{\alpha\sqrt{p}\ub_i[k]^\H\Urf^\H \tilde{\hb}_i[k] \tilde{s}_i[k]}_{\text{desired signal}} \nonumber\\
       &  \ + \underbrace{\alpha\sqrt{p}\sum\limits_{j\neq i} \ub_i[k]^\H\Urf^\H \tilde{\hb}_j[k]\tilde{s}_j[k]}_{\text{interference}} + \underbrace{\ub_i[k]^\H\eb[k]}_{   \text{AWGN and QD} }.
    \end{align}

\vspace{2mm}
\section{Hybrid Receiver Design}
\subsection{Problem Formulation}
Based on \eqref{eq:combined signal model}, the signal-to-interference-plus-noise-and-distortion ratio (SINDR) of the $i$-th UE at the $k$-th subcarrier can be expressed as
\begin{equation}
    \zeta_i[k] = \frac{p \alpha^2 |\ub_i[k]^\H\Urf^\H\tilde{\hb}_i[k]|^2 }{p \alpha^2 \sum_{j\neq i} |\ub_i[k]^\H\Urf^\H\tilde{\hb}_j[k]|^2+ \ub_i[k]^\H\Cb_{\eb_k}\ub_i[k]},
\end{equation}
with $\Cb_{\eb_k} = \Es\left[\eb[k]\eb[k]^\H\right]=\Cb_{\tilde{\etab}_k}+\alpha^2\sigman^2 \Urf^\H\Urf$ and where $\Cb_{\tilde{\etab}_k} = \Es\left[\tilde{\etab}[k]\tilde{\etab} [k]^\H\right]$ denotes the frequency-domain covariance matrix of the QD. Treating the interference-plus-noise-and-distortion term as a Gaussian random variable, we obtain the SE as \cite{fan2015uplink}
\begin{equation}\label{eq:achievable sum rate}
	R=  \frac{1}{K}\sum\limits_{k=1}^K \sum\limits_{i=1}^I  \log_2\left( 1+ \zeta_i[k]\right).
\end{equation}
We observe that $\Cb_{\tilde{\etab}_k}$ is required to compute the SE in \eqref{eq:achievable sum rate}. However, obtaining $\Cb_{\tilde{\etab}_k}$ is challenging because both the temporal and spatial correlation of the QD must be considered. We herein present a closed-form approximation in the following proposition.

\vspace{-5pt}
\begin{proposition}\label{prop:Cov_eta_fd}
Let $\diag(\Ab)$ denote a diagonal matrix with the same diagonal entries as $\Ab$. $\Cb_{\tilde{\etab}_k} $ can be approximated as
	\begin{align}\label{eq:quantization noise in frequency domain}
		\Cb_{\tilde{\etab}_k} & \approx  \gamma(1-\gamma)   \frac{p}{\Nc}\sum\limits_{k=0}^{K-1} \diag(\Urf^\H \tilde{\Hb}[k]\tilde{\Hb}[k]^\H\Urf) \nonumber\\
  & \phantom{=} \ +\gamma(1-\gamma) \sigman^2\Urf^\H\Urf,
	\end{align}
where its accuracy increases with the ADC resolution.
\end{proposition}

\noindent Proposition~\ref{prop:Cov_eta_fd} can be obtained through the DFT of the time-domain covariance matrices $\Cb_{\rb}[\iota] = \Es\left[ \rb[n]\rb[n-\iota]^\H\right]$ and $\Cb_{\etab}[\iota] = \Es\left[ \etab[n]\etab[n-\iota]^\H\right]$ as well as the diagonal approximation $\Cb_{\etab}[0]\approx \gamma(1-\gamma)\diag(\Urf^\H\Cb_{\rb}[0]\Urf)$ \cite{ma2024joint}. The detailed proof is omitted due to the space limitations. Note that $\Cb_{\etab}[\iota]$ includes both the temporal and spatial correlation of the QD.

Define $\Hb_{\rm e} =  \frac{1}{K}\sum_{k=0}^{K-1} \diag(\Urf^\H \tilde{\Hb}[k]\tilde{\Hb}[k]^\H \Urf)$. From \eqref{eq:quantization noise in frequency domain},  we obtain the SINDR as
\begin{equation}\label{eq:effective snr}
    \zeta_i[k]\approx \frac{ |\ub_i[k]^\H\Urf^\H\tilde{\hb}_i[k]|^2 }{\sum_{j\neq i} |\ub_i[k]^\H\Urf^\H\tilde{\hb}_j[k]|^2+ \ub_i[k]^\H\Cb_{\eb}\ub_i[k]},
    \vspace{6pt}
\end{equation}
with
\begin{equation}\label{eq:effective noise}
    \Cb_{\eb} = \frac{\gamma}{(1-\gamma)\beta}\Hb_{\rm e} + \frac{1}{\rho(1-\gamma)}\Urf^\H\Urf,
\end{equation}
where $\rho = p/\sigman^2 $ is the SNR.
We note that $\Cb_{\eb}$ represents the covariance of the overall effective noise of the system, which is jointly affected by the OSR, the ADC resolution, and the SNR. 

Building on \eqref{eq:effective noise}, we aim to jointly optimize the analog and digital combiners to maximize the SE. Specifically, the problem of interest is formulated as
\begin{subequations}\label{prob:HC design}
    \begin{align}
            & \underset{\Urf,\{\Ubb[k]\}_{k=1}^K}{\textrm{maximize}} \quad R \label{obj:rate} \\
            & \hspace{5mm} \textrm{subject to} \hspace{4mm}  \quad |\Urf(m,n)|=\frac{1}{\sqrt{M}},~\forall m,n, \label{const:analog combiner}
    \end{align}
\end{subequations}
where $\Urf(m,n)$ is the entry of $\Urf$ in the $m$-th row and $n$-th column. Due to the hardware constraint of the analog combiner and the non-convex objective function, solving \eqref{prob:HC design} poses a considerable challenge. We devise an efficient solution to \eqref{prob:HC design} based on fractional programming \cite{shen2018fractional} below.

\subsection{Solution}
We first show that \eqref{prob:HC design} can be solved through an equivalent but more tractable problem in the following proposition.

\begin{proposition}
\label{prop:FP equivalent expression}
The objective function in \eqref{obj:rate} is equivalent to the maximization of
\begin{align}\label{eq:fq}
    &f_q\left(\Urf, \{ \Ubb [k], \tb[k], \qb[k]\}_{k=1}^K\right)\nonumber \\
    &= \frac{G}{K\ln2} +\frac{1}{K} \sum_{k,i} \log_2(1+t_i[k]) -\frac{1}{K\ln 2} \sum_{k,i} t_i[k]
\end{align}
with respect to $\{ \tb[k], \qb[k]\}_{k=1}^K$, where $\tb[k]\in \Rs^{I}$ and $\qb[k] \in \Cs^{I},~\forall k$ are auxiliary variables and 
\begin{align}\label{eq:G}
   &G = \sum_{k,i}2\sqrt{(t_i[k]+1)}\Re\{q_i[k]^* \ub_i[k]^\H\Urf^\H\tilde{\hb}_i[k]\} \nonumber \\
   &  -\sum_{k,i} |q_i[k]|^2 \left( \sum\limits_{j=1}^I |\ub_i[k]^\H\Urf^\H\tilde{\hb}_j[k]|^2+ \ub_i[k]^\H\Cb_{\eb}\ub_i[k] \right),
\end{align}
where $t_i[k]$ and $q_i[k]$ are the $i$-th elements of $\tb[k]$ and $\qb[k]$, respectively, and $(\cdot)^*$ represents the complex conjugate. The optimal $t_i[k]$ and $q_i[k]$ are given by
   \begingroup
\allowdisplaybreaks
    \begin{align}
        t_i^{\star} [k] & = \frac{ |\ub_i[k]^\H\Urf^\H\tilde{\hb}_i[k]|^2 }{\sum_{j\neq i} |\ub_i[k]^\H\Urf^\H\tilde{\hb}_j[k]|^2+ \ub_i[k]^\H\Cb_{\eb}\ub_i[k]},\label{eq:optimal rmk} \\
        q_i^{\star} [k]& =\frac{\sqrt{(t_i[k]+1)}\ub_i[k]^\H\Urf^\H\tilde{\hb}_i[k]}{ \sum_{j=1}^I |\ub_i[k]^\H\Urf^\H\tilde{\hb}_j[k]|^2+ \ub_i[k]^\H\Cb_{\eb}\ub_i[k]}. \label{eq:optimal qmk}
    \end{align}
\endgroup
\end{proposition}

\noindent The detailed proof is omitted due to the space limitations. Proposition \ref{prop:FP equivalent expression} shows that the solutions to \eqref{prob:HC design} can be obtained as
\begin{align}\label{pb:alternative form in Multiuser cases}
   & \{ \Urf^{\star}, \{ \Ubb^{\star}[k],\tb^{\star}[k], \qb^{\star}[k]\}_{k=1}^K\} \nonumber \\
    & = \ \arg\max \quad   f_q(\Urf, \{ \Ubb [k], \tb[k], \qb[k]\}_{k=1}^K) \nonumber \\
    & \ \phantom{=} \ \text{subject to} \quad \eqref{const:analog combiner} .
\end{align}

Let $\{\Ubb[k]\}$, $\{\tb[k]\}$ and $\{\qb[k]\}$ represent $\{\Ubb[k]\}_{k=1}^K$, $\{\tb[k]\}_{k=1}^K$ and $\{\qb[k]\}_{k=1}^K$ for brevity. We observe that  $\Urf$, $\{\Ubb[k]\}$, $\{\tb[k]\}$, and $\{\qb[k]\}$ can be iteratively updated to solve \eqref{pb:alternative form in Multiuser cases}. Since the optimal $\{\tb[k]\}$ and $\{\qb[k]\}$ are given in \eqref{eq:optimal rmk} and \eqref{eq:optimal qmk}, respectively, we aim to derive the optimal $\Urf$ and $\{\Ubb[k]\}$ in the following.

\smallskip

\textit{Analog combiner design:} Define $\underline{\Hb} = \frac{1}{K}\sum_{k=0}^{K-1} \tilde{\Hb}[k]\tilde{\Hb}[k]^\H$, $\Yb[k] =  \sum_{i=1}^{I} \tilde{\hb}_i[k]\tilde{\hb}_i[k]^\H,~\forall k$,~and
\begin{align}
   \Xb  &= \sum_{k,i}\sqrt{(t_i[k]+1)}q_i[k]^* \tilde{\hb}_i[k] \ub_{i}[k]^\H, \\[-3pt]
\Zb[k]  &=  \sum\limits_{i=1}^{I}  |q_i[k]|^2\ub_i[k] \ub_i[k]^\H,~\forall k.
\end{align}
The objective function with respect to $\Urf$ is expressed as
    \begin{align}
    g\left( \Ub_{\rm RF}\right) & = 2\Re\{\tr \left(\Xb\Urf^\H\right)\}- \sum\limits_{k=1}^K\tr\left( \Zb[k]\Urf^\H \Yb[k]\Urf\right) \nonumber \\[-5pt]
    & \phantom{=} \  -\frac{\gamma}{(1-\gamma)\beta}\sum\limits_{k=1}^K\tr\left( \diag(\Zb[k])\Urf^\H\underline{\Hb}\Urf \right) \nonumber \\[-2pt]
    & \phantom{=} \ -  \frac{1}{\rho(1-\gamma)}\sum\limits_{k=1}^K\tr\left(\Zb[k]\Urf^\H\Urf \right).
\end{align}
With the other variables fixed, the design of the analog combiner $\Urf$ can be expressed as
\begin{align}\label{pb:analog BF design MU}
\begin{split}
        & \underset{\Urf}{\mathrm{maximize}} \quad g\left( \Urf\right) \\
        & \, \text{subject to} \quad \eqref{const:analog combiner}.
\end{split}
\end{align}
We adopt the projected gradient ascent (PGA) algorithm \cite{ma2024switch} to efficiently solve \eqref{pb:analog BF design MU} in an iterative fashion, where the gradient of $g(\Frf)$ is given by
\begin{align}
     &\nabla_{\Urf} g\left( \Urf\right)=2\Xb -\frac{2}{\rho(1-\gamma)}\sum\limits_{k=1}^K\Urf \Zb[k] \nonumber  \\[-3pt]
     &-\frac{2\gamma}{(1-\gamma)\beta}\sum\limits_{k=1}^K \underline{\Hb}\Urf \diag(\Zb[k])-  2\sum\limits_{k=1}^K\Yb[k]\Urf\Zb[k]. \label{eq:PGA gradient P2P}
\end{align}
The detailed procedure is summarized in Algorithm~\ref{alg:PGA algorithm}. In step 4, the normalized gradient $\nabla \Urf$ is obtained and employed later in step 6, where the step size $\mu$ is determined via backtracking line search as in step 5 and $[\cdot]_{\Ucl}$ denotes the projection onto $\Ucl= \{\Urf: |\Urf(m,n)|=\frac{1}{\sqrt{M}},~\forall m,n\}$. The iterations continue until the stopping criterion is satisfied. 

\begin{algorithm}[t!]
\small
\caption{ PGA algorithm for solving \eqref{pb:analog BF design MU}}\label{alg:PGA algorithm}
\LinesNumbered 
\KwOut{$\Ub_{\rm RF}^{\star}$}
Initialize $\nu \in (0,0.5)$, $ c \in (0,1)$, $\Urf\in \Ucl$, and $\epsilon$\;
\Repeat{$  |g(\Urf')-g(\Urf)| \leq \epsilon$ }{
$\Urf'\leftarrow \Urf$.

  Obtain $\nabla \Urf =\frac{\nabla_{\Urf}g(\Urf)}{\left\| \nabla_{\Urf}g(\Urf)\right\|_F}$ as in \eqref{eq:PGA gradient P2P}.

 Start with $\mu=1$, repeat $\mu \leftarrow  c \mu$ until
 $g\left( [\Urf+\mu\nabla \Urf]_{\Ucl}\right)> g\left( \Urf\right)+ \nu \mu \|\nabla\Urf\|_F^2$.

 $\Urf \leftarrow [\Urf+\mu\nabla \Urf]_{\Ucl}$.
 }
\end{algorithm}

\smallskip

\textit{Digital combiner design:} For given $\Urf$ and $\{\tb[k], \qb[k]\}_{k=1}^K$, the optimal digital combiner is obtained as
\begin{equation}\label{pb:digital BF design MU}
    \{ \Ubb^{\star}[k]\}_{k=1}^K =\arg\max \; G ,
\end{equation}
where $G$ is given in \eqref{eq:G}.
From the first-order optimality condition, we can derive the optimal digital combiner for the $i$-th UE at the $k$-th subcarrier as
\begin{equation}\label{eq:CF digital BF MU}
    \ub_{i}^{\star}[k] =  \xi_i[k] \left(\Urf^\H\Yb[k]\Urf+  \Cb_{\eb} \right)^{-1} \Urf^\H \tilde{\hb}_i[k] ,~\forall i,k,
\end{equation}
with $\xi_i[k] = \frac{q_i[k]^* \sqrt{t_i[k]+1}}{|q_i[k]|^2} $.

\smallskip

The proposed hybrid receiver design for solving \eqref{prob:HC design} is summarized in Algorithm~\ref{alg:FP relaxed HBF design}, which can be initialized based on the receiver design in \cite{zhu2017low}. Specifically, let $\underline{\Hb} =\Vb \Sigmab \Vb^\H$ be the singular value decomposition (SVD) of $\underline{\Hb}$, where $\Vb=[\vb_1,\ldots,\vb_{M}]$ is a unitary matrix and $\Sigmab$ is a diagonal matrix with the singular values on the main diagonal in descending order. Then, $\Urf$ is initialized as 
\begin{equation}\label{eq:SVD analog combiner}
    \Urf=\frac{1}{\sqrt{M}}e^{j\angle \hat{\Vb} },
\end{equation}
with $\hat{\Vb}=[\vb_1,\ldots,\vb_{\Nrf}]$ and where $\angle \hat{\Vb}$ denotes the phase of the entries of $\hat{\Vb}$. The digital combiner can be initialized based on the MMSE criterion. Specifically, we have 
\begin{equation}\label{eq:mmse digital combiner}
    \ub_i[k]=\alpha \sqrt{p}\Jb[k]^{-1}\Urf^\H\tilde{\hb}_i[k],~\forall i,k,
\end{equation}
with $\Jb[k]= \alpha^2 p \big(\sum_{i=1}^I \Urf^\H \tilde{\hb}_i[k]\tilde{\hb}_i[k]^\H \Urf + \Cb_{\eb} \big)$.

We note that Algorithm \ref{alg:FP relaxed HBF design} converges to a stationary point since the objective function \eqref{eq:fq} is monotonically nondecreasing over the iterations \cite{shen2018fractional}. Let $N_{\rm pga}$ and $N_{\rm iter}$ denote the number of iterations of Algorithms~\ref{alg:PGA algorithm} and~\ref{alg:FP relaxed HBF design}, respectively. When $I\leq\Nrf$ and $\Nrf \ll M$, the computational complexities of Algorithms~\ref{alg:PGA algorithm} and~\ref{alg:FP relaxed HBF design} are in the order of $\Ocl(4N_{\rm pga}KM^2\Nrf)$ and $N_{\rm iter}\Ocl(4N_{\rm pga}KM^2\Nrf+2KM^2\Nrf^2)$, respectively, which are mainly due to the matrix multiplications.
\begin{algorithm}[t!]
\small
\caption{ Hybrid receiver design for solving \eqref{prob:HC design}}\label{alg:FP relaxed HBF design}
\LinesNumbered 
\KwOut{$\Ub_{\rm RF}^{\star}, \{ \Ubb[k]^{\star}\}_{k=1}^K$}
Initialize $\Ub_{\rm RF}\in \Ucl$ and $\{ \Ubb [k]\}_{k=1}^K$.

\Repeat{{\rm a predefined stopping criterion is satisfied}}{
  Update $\{ \tb^{\star}[k]\}_{k=1}^K $ as in \eqref{eq:optimal rmk}.

 Update $\{ \qb^{\star}[k]\}_{k=1}^K $ as in \eqref{eq:optimal qmk}.

Update $\Ub_{\rm RF}^{\star}$ via Algorithm~\ref{alg:PGA algorithm}.

  Update $ \{ \Ubb^{\star}[k]\}_{k=1}^K$ as in \eqref{eq:CF digital BF MU}.
 }

 \end{algorithm}

\begin{figure*}[t!]
    \hspace{-1mm}
    \subfigure[]
        {\label{fig:SE vs SNR}\includegraphics[scale=0.41]{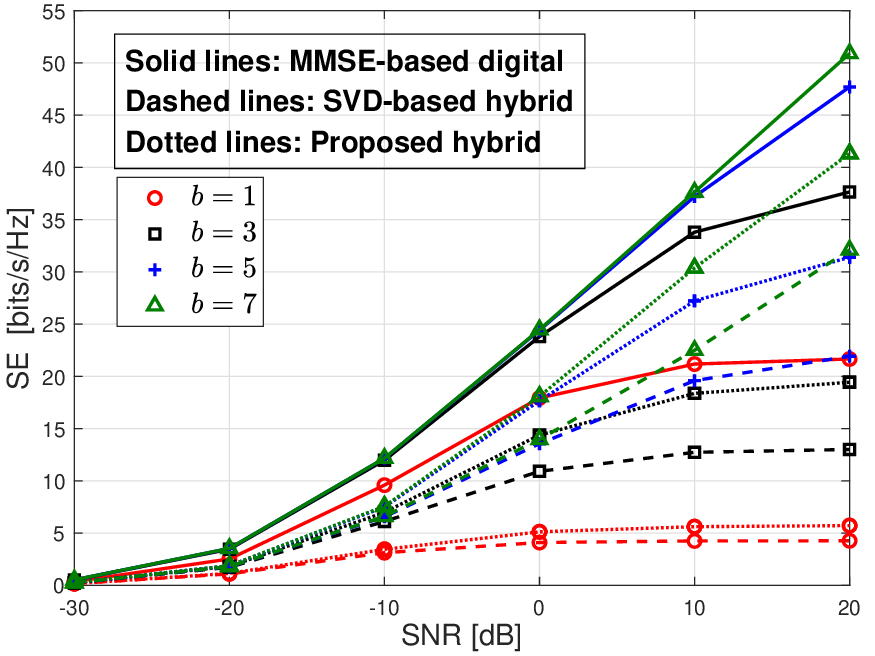}}
    \hspace{-3mm}
    \subfigure[]
        {\label{fig:SE vs bandwidth} \includegraphics[scale=0.41]{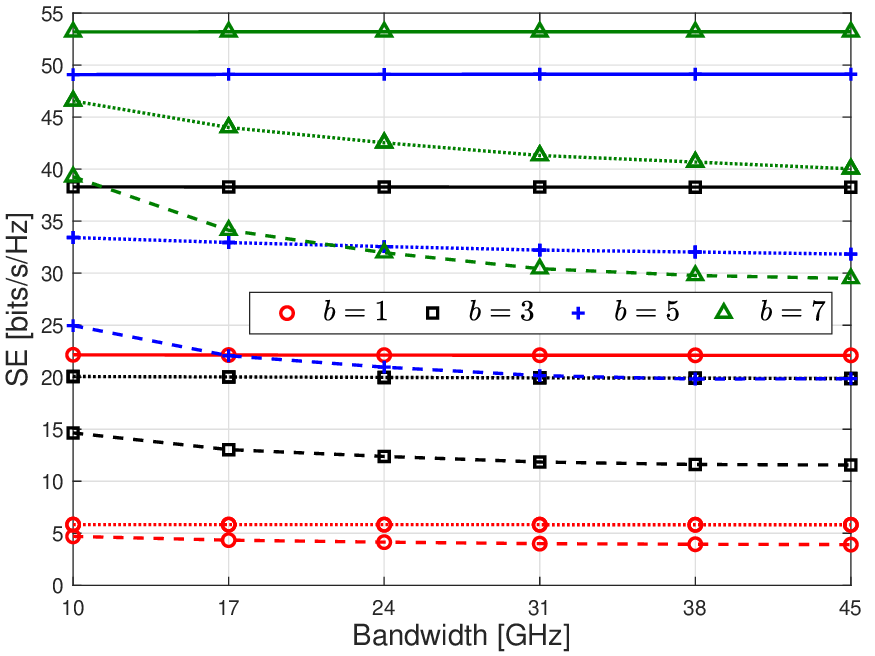}}
    \hspace{-3mm}
    \subfigure[  ]
        {\label{fig:SNR vs OSR} \includegraphics[scale=0.41]{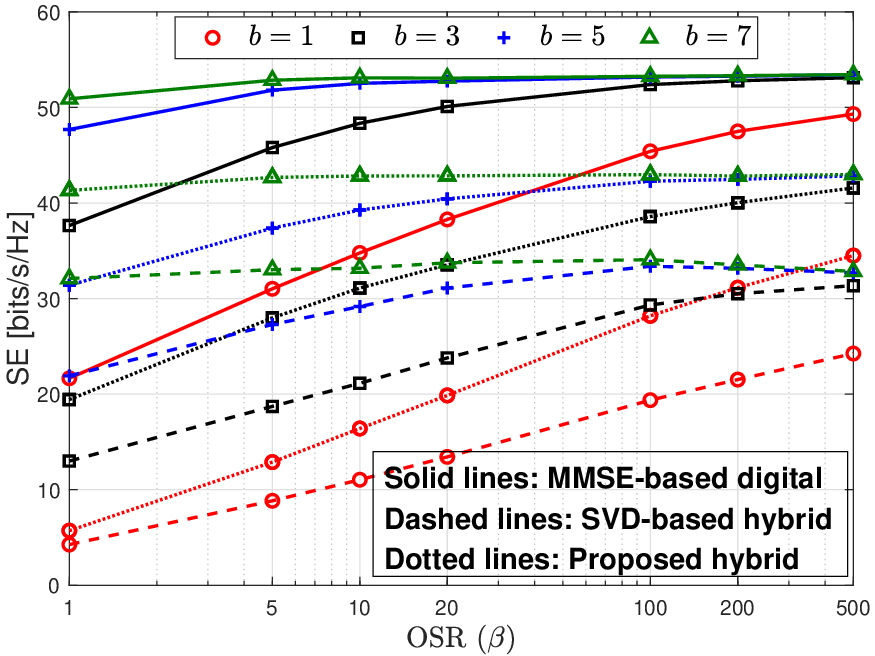}}
    \vspace{-2mm}
    \caption{SE versus SNR, bandwidth, and OSR with $M=128$ and $\Nrf=I=4$. (a): SE versus SNR ($\rho$) with $\beta=1$ and $B_{\rm w}=12.8~\ghz$; (b): SE versus bandwidth ($B_{\rm w}$) with $\beta=1$ and $\SNR=20~\dB$; (c): SE versus OSR ($\beta$) with $\SNR=20~\dB$ and $B_{\rm w}=12.8~\ghz$. In the three figures, the groups of solid, dashed, and dotted lines represent the MMSE-based fully digital, SVD-based hybrid, and proposed hybrid receiver designs, respectively.}
    \label{fig:Sum rate performance}
    \vspace{-3mm}
\end{figure*}

\section{Numerical Results}\label{sec:simulations}

In the simulations, 
the channel vector between the $i$-th UE and the BS at the $k$-th subcarrier is modeled as \cite{ma2023beam}
\begin{equation}
    \tilde{\hb}_i[k]=\sqrt{\frac{M}{L}}\sum\limits_{\ell=1}^{L}g_{i,\ell} [k]\ab(\theta_{i,\ell},f_k),
\end{equation}
with $g_{i,\ell}[k]= \sum_{d=0}^{D-1}\lambda_{i,\ell} p_{\rm rc}(\frac{d}{\fs}-\tau_{i,\ell})e^{-j\frac{2\pi k}{\Nc}d} $ and where $\lambda_{i,\ell}$, $\tau_{i,\ell}$, and $\theta_{i,\ell} $ are the $\ell$-th gain, delay, and angle of arrival (AoA), respectively, for each of the $L$ paths. In the simulations, the pulse-shaping filter $ p_{\rm rc}(t) $ is modeled by the raised-cosine function with roll-off factor of $1$ \cite{alkhateeb2016frequency}. We note that the roll-off factor only affects the channel and therefore has a minor impact on the performance comparison. We assume $L=3$ paths, $\lambda_{i,\ell}\sim\Ccl\Ncl(0,1)$, $\tau_{i,\ell}\sim \Dcl \big[0,\frac{D_0}{\Bw} \big]$ (with $D_0=K/4$ as in \cite{ma2023analysis}), and $\theta_{i,\ell}\sim \Dcl[0,2\pi]$, where $\Dcl[a,b]$ represents the uniform distribution in the interval $[a, b]$.
The array steering vector is expressed as $\ab(\theta_{i,\ell},f_k)=\frac{1}{\sqrt{M}}[1,e^{-j\pi\frac{f_k}{\fc} \sin(\theta_{i,\ell})},\ldots, e^{-j(M-1)\pi \frac{f_k}{\fc}\sin(\theta_{i,\ell})}]^\T$. Further- \linebreak more, we set $\fc=300~\ghz$ and $K=128$ subcarriers. The following results are obtained by averaging over $10^3$ independent channel realizations.

Fig.~\ref{fig:Sum rate performance} shows the SE versus the SNR, the bandwidth, and the OSR with $M=128$ and $\Nrf=I=4$. Here, ``MMSE-based digital'', ``SVD-based hybrid'', and ``Proposed hybrid'' represent the MMSE-based fully digital receiver design, the hybrid receiver design based on \eqref{eq:SVD analog combiner}--\eqref{eq:mmse digital combiner}, and the proposed hybrid receiver design, respectively. We observe from Fig.~\ref{fig:SE vs SNR} that the proposed hybrid receiver design significantly outperforms the SVD-based hybrid receiver design, especially at high SNRs. For instance, with $b=3$, the former provides a gain of $43\%$ over the latter at $\SNR=20~\dB$. However, the hybrid receivers exhibit large SE gaps compared with the MMSE-based fully digital receiver, especially for low-resolution cases. With $b=7$, the proposed hybrid receiver design attains $80\%$ of the SE achieved by the MMSE-based fully digital receiver design. In contrast, the former only achieves $26\%$ SE of the latter with $b=1$. Additionally, we observe from Fig.~\ref{fig:SE vs bandwidth} that the SE of the hybrid receivers degrades with increasing bandwidth due to the beam squint effect, aligned with the observations in \cite{dai2022delay,ma2023beam,ma2024switch}. Remarkably, the SE of the proposed hybrid receiver design is less affected by the increase in bandwidth for lower resolution systems. For example, when the bandwidth increases from $10~\ghz$ to $45~\ghz$, the SE of the proposed scheme decreases by $6.5~\text{bit/s/Hz}$ with $b=7$ and only by $1.6~\text{bit/s/Hz}$ with $b=5$. This is likely because the QD overwhelms the array gain loss due to squinted beams in low-resolution systems.

Fig.~\ref{fig:SNR vs OSR} shows the SE versus the OSR with $\SNR=20~\dB$ and $B_{\rm w}=12.8~\ghz$. It is shown that increasing the OSR can significantly enhance the SE of both the hybrid and fully digital receivers, especially with low-resolution ADCs. For instance, with $b=1$, a $126\%$ SE improvement is achieved when the OSR increases from $1$ to $5$ compared with $44\%$ with $b=3$.  Such results align with the findings in \cite{ma2023analysis}. When the OSR is sufficiently large, the QD can be sufficiently suppressed (as seen from \eqref{eq:effective noise}), leading to an SE similar to that obtained with the fully digital or hybrid receivers with $b \in \{3, 5, 7\}$ at $\beta=500$. However, increasing the OSR also results in a higher power consumption of the ADCs. This is because a $b$-bit ADC typically has a power consumption of $P_{\rm ADC}=\kappa f_{\rm s}2^b$ \cite{ma2023analysis} where $\kappa$ represents the figure of merit (FoM). Hence, the OSR requires to be judiciously chosen to achieve a satisfactory trade-off between SE and EE. 

\begin{figure}[t!]
        \centering
        \includegraphics[scale=0.5]{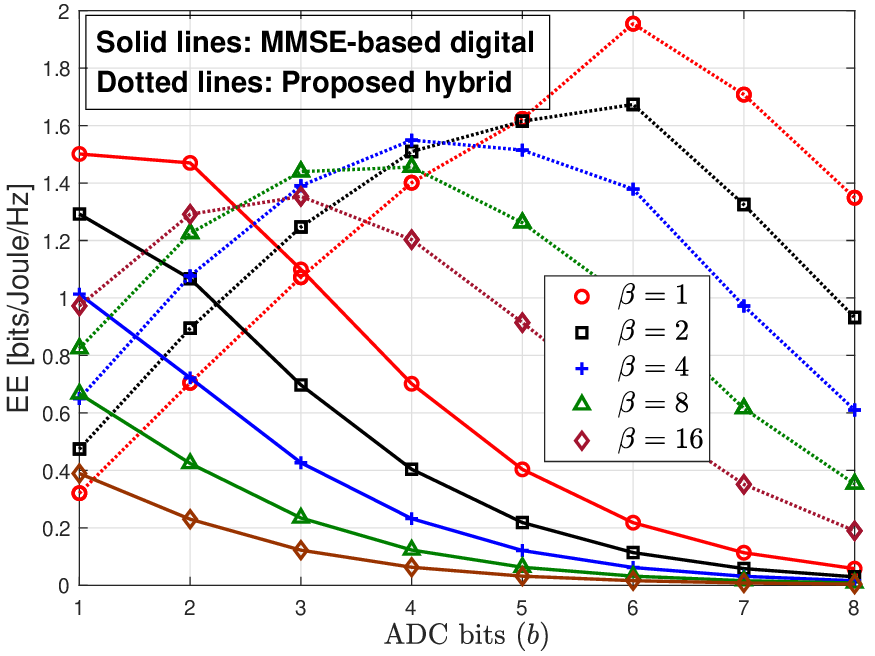}
        \caption{EE versus ADC bits and OSR with $M=128$, $\Nrf=I=4$, $\SNR=20~\dB$, and $B_{\rm w}=12.8~\ghz$.}
        \label{fig:EE versus bit and OSR}
    \end{figure}

Next, we evaluate the EE, which is defined as the ratio between the SE and the total power consumption of the receiver \cite{abbas2017millimeter}. The total power consumption of the fully digital receiver and the considered hybrid receiver are given by $P_{\rm D}=  M\left( P_{\rm LNA} + P_{\rm RF}+ 2P_{\rm ADC} \right)$ and $P_{\rm H} =M \left(P_{\rm LNA}+ P_{\rm SP}+\Nrf P_{\rm PS}\right) +\Nrf \left( P_{\rm RF}+P_{\rm C}+2P_{\rm ADC}\right) $, respectively, where $P_{\rm LNA}$, $P_{\rm RF}$, $P_{\rm ADC}$, $P_{\rm SP}$, $P_{\rm C}$, and $P_{\rm PS}$ represent the power consumption of the low-noise amplifier (LNA), RF chain, ADC, splitter, combiner, and phase shifter, respectively. In the following simulations, we set $P_{\rm RF}=43~\mW$, $P_{\rm LNA}=25~\mW$, $P_{\rm SP}$/$P_{\rm C}=19.5~\mW$, $P_{\rm PS}=23~\mW$, and $\kappa=494~{\rm fJ/step/Hz}$ \cite{abbas2017millimeter}. Fig.~\ref{fig:EE versus bit and OSR} shows the EE of the MMSE-based fully digital receiver design and the proposed hybrid receiver design versus the ADC bits for different OSRs with $M=128$, $\Nrf=I=4$, $\SNR=20~\dB$, and $B_{\rm w}=12.8~\ghz$. We make the following observations. First, the EE of the fully digital receiver decreases as the ADC bits and OSR increase, while that of the hybrid receiver achieves the highest EE with $b\in\{3,4,6\}$ and $\beta\in \{1,2,4,8,16\}$. This is because the large number of RF chains in the fully digital receiver entails excessively high power consumption, which offsets the resulting SE improvements due to the higher ADC resolution and OSR. In contrast, the number of RF chains in the hybrid receiver is significantly smaller than in the fully digital case, making the increase in ADC bits beneficial in terms of EE for relatively low ADC resolution. Second, the OSR with $b\in \{1,2\}$ in the hybrid receiver can significantly improve the EE, while it reduces the EE with $b\geq 5$. This is because increasing the OSR for higher resolution ADCs contributes only to minor SE improvements but results in significantly higher power consumption of the ADCs. Third, from the EE maximization perspective, the fully digital receiver with $b=1$ or $b=2$ and without oversampling can achieve better performance than the hybrid receiver. In contrast, the hybrid receiver with $b>3$ and oversampling performs significantly better than the fully digital receiver. Moreover, the results presented in Figs.~\ref{fig:Sum rate performance} and \ref{fig:EE versus bit and OSR} indicate the potential of joint beamforming design and bit allocation in improving the SE--EE trade-off, as demonstrated in \cite{ma2024joint}.

\section{Conclusion}
We studied the hybrid receiver design in a massive MIMO-OFDM system with low-resolution ADCs and oversampling. Specifically, we first derived a closed-form approximation for the frequency-domain covariance matrix of the QD, which facilitates the evaluation of SE. Then we proposed an efficient algorithm to jointly optimize the analog and digital combiners to maximize the SE. Numerical simulations validated the superiority of the proposed algorithm over the considered benchmarks. Furthermore, the results showed that the hybrid receiver is less affected by the beam squint effect with low-resolution ADCs. In addition, the EE comparison showed that the proposed hybrid receiver design with oversampling significantly outperforms the considered fully digital receiver when slightly more (larger than $3$) ADC bits are used per RF chain.

\section*{Acknowledgements}

This work was supported by the Research Council of Finland (332362 EERA, 336449 Profi6, 348396 HIGH-6G, 357504 EETCAMD, 354901 DIRECTION, and 369116 6G~Flagship) and by the European Commission (101095759 Hexa-X-II).

\bibliographystyle{IEEEtran}
\bibliography{conf_short,jour_short,refs-my}

\end{document}